\documentclass[epj,final]{svjour}

\usepackage{latexsym}
\usepackage{url}
\usepackage{amsfonts}
\usepackage{amsmath, amssymb}
\usepackage{color}
\RequirePackage{graphicx, subcaption}

\begin{document}
\title{Smearing of primordial gravitational waves}
\author{M.~Samsonyan\inst{1}, A.A.~Kocharyan\inst{2}, V.G.~Gurzadyan\inst{1,3,4}
}                     
%
%
\institute{Center for Cosmology and Astrophysics, Alikhanian National Laboratory and Yerevan State University, Yerevan, Armenia  \and 
School of Physics and Astronomy, Monash University, Clayton, Australia \and
SIA, Sapienza Universita di Roma, Rome, Italy \and  e-mail: gurzadyan@yerphi.am (corresponding author)}
\date{Received: date / Revised version: date}
%

\abstract{A mechanism for smearing of the primordial gravitational waves during the radiation-dominated phase of the evolution of the Universe is considered. It is shown that the primordial gravitational waves can possess hyperbolicity features due to their propagation through the matter inhomogeneities. This mechanism of smearing can lead to the flattening of the original gravitational wave spectrum and hence has to be taken into account at the interpretation of the properties of primordial gravitational background on the detection of which are oriented ongoing and forthcoming experimental facilities.} 
%
\maketitle

\section{Introduction}

Primordial gravitational waves are considered as potentially unique tools for the study of the very early phases of the Universe, including the features of the inflationary models \cite{St1,St2}. The studies of theoretical models of production of primordial gravitational waves, of their basic properties, as well as the prospects of their experimental detection were essentially broadened upon the ongoing measurements by LIGO-Virgo-Kagra collaboration, NANOGrav consortium  \cite{Kam,Koh,Wang,Bal,Chang,Yu} and references therein. New possibilities regarding the study of primordial gravitational waves can be provided by forthcoming LISA interferometer, e.g. \cite{Ric}. It is outlined that, primordial gravitational wave background, when detected, can open an entirely new window to a broad scope of effects, as did the detection of the electromagnetic primordial waves, the Cosmic Microwave Background (CMB). Indeed, the latter enabled probing not only the effects at the last scattering surface but also before and after, including the features of the large scale matter distribution, cosmic voids, clusters of galaxies via gravitational lensing, etc  \cite{Mukhanov,Holz}.  

It is believed that, first, the primordial gravitational waves have to be originated from quantum fluctuations \cite{St1}, second, then had freely propagated during the subsequent expansion of the Universe, largely keeping their original features. Therefore their features, when detected, are expected to provide a direct insight to the inflationary phase,  on the involved scalar fields, on the particle production, etc, for details, see \cite{Cap}.
   
Below, we consider a tiny effect which nevertheless can influence the study and interpretation of the properties of the primordial gravitational waves, since this effect provides a principal possibility of the smearing of those properties at radiation-dominated phase of the evolution of the Universe. Namely, we study the propagation of gravitational waves in that phase of the Universe and reveal the conditions for their instability (hyperbolicity) due to matter inhomogeneities which can smear their primordial signatures.  This mechanism is dealing with the geodesic flows as formulated in the theory of dynamical systems \cite{An,Arn}, and, for example, enabled to reveal the hyperbolicity of the photon beams induced at their propagation within cosmic voids \cite{GCMB,GK1,GK2}. The hyperbolicity of geodesic flows was instrumental to conclude regarding the void nature of the Cold Spot \cite{spot}, a non-Gaussian anomaly in the CMB sky, by means of the estimation of Kolmogorov stochasticity parameter and using the Planck satellite's data.  Hyperbolicity of photon beams depending on the parameters of the cosmic voids and their redshifts and its role in the analysis of the high redshift galactic surveys is studied in \cite{Sam1,Sam2,Sam3}.

Another aspect of the importance of the detection and accurate interpretation of the gravitational wave background besides addressing the early cosmos and high energy physics scale is also expected in the probing of modified gravity theories \cite{Capo,GS} and other approaches regarding the cosmological (observational) tensions \cite{Val,Hu,Ri}. Note, that, NANOGrav and collaboration data \cite{Ar,Ag} on the nano-hertz gravitational wave background are already used to constrain modified gravity theories, (see, e.g. \cite{Can} and references therein), thus complementing the other means of high accuracy testing of General Relativity \cite{ZNP,Kop,C1,C2}.

\section{The perturbed metric}
     
We consider scalarly perturbed Robertson-Walker metric \cite{Holz}
\begin{equation}
ds^2=-(1+2 \phi )d\tau^2+(1-2\phi)a^2(\tau)\left[\frac{dr^2}{1-kr^2}+r^2(d\theta^2+\sin^2\theta d\varphi^2)\right]\label{metric},
\end{equation}
where $\phi$ depends on all four coordinates $\tau, r, \theta, \varphi$. The matrix form of the metric is
\begin{equation}
g_{\mu\nu}=\left(
\begin{array}{cccc}
 -(1+2 \phi) & 0 & 0 & 0 \\
 0 & \frac{(1-2 \phi) a^2(\tau) }{1-k r^2} & 0 & 0 \\
 0 & 0 & (1-2 \phi)a^2(\tau) r^2 & 0 \\
 0 & 0 & 0 & (1-2 \phi)  a^2(\tau) r^2  \sin ^2\theta  \\
\end{array}
\right)
\end{equation}
The energy-momentum tensor is given by
\begin{equation}
T_{\mu\nu}=(\rho +p)U_\mu U_\nu +p g_{\mu\nu}, \label{Tmunu}
\end{equation}
where $U_\mu$ is  a unit timelike vector $U_\mu=(1,0,0,0)$, $U_\mu^2=-1$. For the metric \eqref{metric} it satisfies
\begin{eqnarray}
&&(U_\mu)^2=-1=U_0U_0 g^{\tau\tau}=(U_0)^2\frac{-1}{1+2\phi}, \\
&&(U^0)^2=1+2\phi, \quad U^0\approx 1+\phi \quad \text{for} \quad  \phi \ll 1.
\end{eqnarray}
Then the non-vanishing components of the energy-momentum tensor for scalarly perturbed Robertson-Walker metric
 are
\begin{eqnarray}
&&T_{\tau\tau}=(\rho+p)U_0U_0+pg_{\tau\tau}=\rho(1+2\phi), \\
&&T_{rr}=pg_{rr}=\frac{p(1-2\phi) a^2(\tau)}{1-kr^2}, \\
&&T_{\theta\theta}=pg_{\theta\theta}=p(1-2\phi)a^2(\tau) r^2,\\
&& T_{\varphi\varphi}=pg_{\varphi\varphi}=p(1-2\phi)a^2(\tau)r^2\sin^2\theta\, .
\end{eqnarray} 
We also can write down non-vanishing components of Einstein tensor for the metric \eqref{metric}  
\begin{eqnarray}
&&G_{\tau\tau}=\frac{3k}{a^2}+3\left(\frac{\dot{a}}{a}\right)^2+2a^{-2}h^{ab}D_aD_b\phi +2a^{-2}\left(6k\phi+\frac{\cot\theta}{r^2}\frac{\partial \phi}{\partial\theta}+\frac{2-3kr^2}{r}\frac{\partial \phi}{\partial r}-3 a\dot{a} \frac{\partial \phi}{\partial \tau}    \right), \label{Gtt}\\
&&G_{rr}=\frac{1}{-1+kr^2}\left(k+\dot{a}^2+2a\ddot{a}-2\left(2 \phi\dot{a}^2+4a\ddot{a}\phi+4a\dot{a}\frac{\partial\phi}{\partial\tau}+a^2\frac{\partial^2\phi}{\partial\tau^2}\right)\right)\label{Grr}\\
&&G_{\tau i}=2\left(\frac{\dot{a}}{a}+\frac{\partial\phi}{\partial\tau} \right)\partial_i \phi, \\
&&G_{ij}=0, \qquad i\neq j, \qquad i,j=r,\theta,\phi ,
\end{eqnarray}
where $D_a$ is the spatial derivative operator. Here we have introduced a notation for the spatial metric 
$h_{ab}=\frac{1}{1-kr^2}dr_adr_b+r^2(d\theta _a d \theta _b+\sin^2\theta d\varphi _a d\varphi _b)$ as in \cite{Holz}.

If we take the same constraints on $\phi$ as in \cite{Holz}, we will have
\begin{eqnarray}
&&G_{\tau\tau}=\frac{3k}{a^2}+3\left(\frac{\dot{a}}{a}\right)^2+2a^{-2}h^{ab}D_aD_b\phi, \\
&&G_{rr}=\frac{k+\dot{a}^2+2a\ddot{a}}{-1+kr^2}.
\end{eqnarray}
The non-vanishing components of the Einstein equation $G_{\mu\nu}+\Lambda g_{\mu\nu}=8\pi GT_{\mu\nu}$ then yield
\begin{eqnarray}
&& 3\frac{\ddot{a}}{a}=\Lambda (1-4\phi)+a^{-2}h^{ab}D_aD_b\phi-4\pi G(\rho+3p+2\phi(\rho-3p)) \label{ddaphi},\\
&& 3\left(\frac{\dot{a}}{a}\right)^2=\Lambda (1+2\phi)-2a^{-2}h^{ab}D_aD_b\phi+8\pi G \rho(1+2\phi)-\frac{3k}{a^2} \label{daphi}.
\end{eqnarray}
This for $p=0$ and $\phi\ll 1$ exactly reproduces the results of \cite{Holz}; obviously, at radiation-dominated phase the role of $\Lambda$ can be neglected. 

\section{Geodesics in presence of scalar perturbations}

The averaged geodesic deviation (Jacobi) equation has the form (see \cite{GK1,Sam1}),
\begin{eqnarray}
\frac{d^2\ell}{d\eta^2}+\frac{\cal R}{d-1}\ell =0, \label{geodeq}
\end{eqnarray}
where $\cal{R}$ is the scalar curvature of space (without conformal factor).
Consider the metric 
\begin{eqnarray}
ds^2=a^2\left[(1+2\phi)d\eta^2-(1-2\phi)\delta_{ij}dx^idx^j\right]\,. \label{metricphi}
\end{eqnarray}
For this metric the spatial scalar curvature takes the form
\begin{eqnarray}
{\cal R}=4 \Delta \phi \qquad   \Delta=\partial_x^2+\partial_y^2+\partial_z^2. \label{Rscalar}
\end{eqnarray}
The gravitational potential $\phi$ satisfies the following equations (see section (7.3.1), Eqs. (7.47)-(7.49) of \cite{Mukhanov})
\begin{eqnarray}
\Delta \phi-3{\cal H}(\phi^\prime+{\cal H} \phi)=4\pi G a^2\overline{\delta\varepsilon}\label{deltaphi},\\
(a\phi)^\prime_{\,\,,i}=4\pi Ga^2 (\varepsilon_0+p_0)\overline{\delta u}_{|| i},\\
\phi^{\prime\prime}+3{\cal H}\phi^\prime+(2{\cal H}^\prime+{\cal H}^2)\phi=4\phi a^2\overline{\delta p}\label{phiprimeprime}.
\end{eqnarray}
Here $\overline{\delta u}_{|| i}$ are the covariant spatial components (gauge-invariant) of the 4-velocity of the fluid element. Combining \eqref{deltaphi} and \eqref{phiprimeprime}, one can write (section (7.3.1), eq. (7.51) of \cite{Mukhanov})
\begin{eqnarray}
\phi^{\prime\prime}+3(1+c_s^2) {\cal H} \phi^{\prime}-c_s^2 \Delta \phi+\left(2{\cal H}^{\prime}+(1+3c_s^2){\cal H}^2\right)\phi =4\pi G a^2 \tau\delta S. \label{phieq}
\end{eqnarray}
The pressure fluctuation is related to the energy density and entropy perturbations 
\begin{eqnarray}
\overline{\delta p}=c_s^2\overline{\delta\varepsilon}+\tau\delta S,\qquad with \qquad
c_s^2\equiv \left(\frac{\partial p}{\partial \varepsilon}\right)_S, \qquad \tau\equiv \left(\frac{\partial p}{\partial S}\right)_\varepsilon.
\end{eqnarray}
Let us consider adiabatic perturbations for which $\delta S=0$, at the domination of relativistic matter with equation of state $p=w\varepsilon$, where $w$ is a positive constant. In this case
\begin{equation}
a\propto \eta^{2/(1+3w)} \qquad c_s^2=w.
\end{equation} 
For plane wave perturbations $\phi=\phi_k(\eta)e^{ik\cdot x}$, eq. \eqref{phieq} becomes
\begin{eqnarray}
\phi_k^{\prime\prime}+\frac{6(1+w)}{1+3w}\frac{1}{\eta} \phi_k^\prime+wk^2 \phi_k=0\, .
\end{eqnarray}
The solution of this equation is given in terms of Bessel functions
\begin{eqnarray}
\phi_k=\eta^{-\nu}\left[C_1J_\nu(\sqrt{w}k\eta)+C_2Y_\nu(\sqrt{w}k\eta)\right], \qquad \nu=\frac{1}{2}\left(\frac{5+3w}{1+3w}\right) \label{phisol}\, ,
\end{eqnarray}
where $C_1$ and $C_2$ are constants.
In a radiation dominated case, when $w=1/3$, the order of Bessel function is $\nu=3/2$, and the solution \eqref{phisol} can be expressed in terms of elementary functions
\begin{eqnarray}
\phi_k=\frac{1}{x^2}\left(C_1\left[\frac{\sin x}{x}-\cos x \right]+C_2\left[\frac{\cos x}{x}+\sin x \right]\right)\, , \label{phiksol}
\end{eqnarray}
where $x=k\eta/\sqrt{3}$.
We can solve eq.\eqref{geodeq} for the above $\phi_k$  and find the distortion of the flow, which will determine the geodesics behavior in this perturbed background. 
For the plane wave the Ricci scalar \eqref{Rscalar} becomes
\begin{eqnarray}
{\cal R}=-4k^2\phi,
\end{eqnarray}
and since $\Delta\phi=-k^2\phi$, the averaged geodesics equation \eqref{geodeq} becomes
\begin{eqnarray}
\frac{d^2\ell}{d\eta^2}-2 k^2\phi\, \ell=0. \label{geoddev}
\end{eqnarray}
Note that, $a\propto \eta^{2/(1+3w)} =a_0\,  \eta$ for $w=1/3$ with constant $a_0$. 
For $x\ll 1$ \eqref{geoddev} becomes
\begin{eqnarray}
\frac{d^2\ell}{d\eta^2}-\frac{C}{\eta^3} \ell=0 \label{geoddev1}\, ,
\end{eqnarray}
where $C=\frac{6\sqrt{3}C_2}{k}$.

\subsection{$C>0$ case}
\begin{figure}
\centering
\includegraphics[scale=0.4]{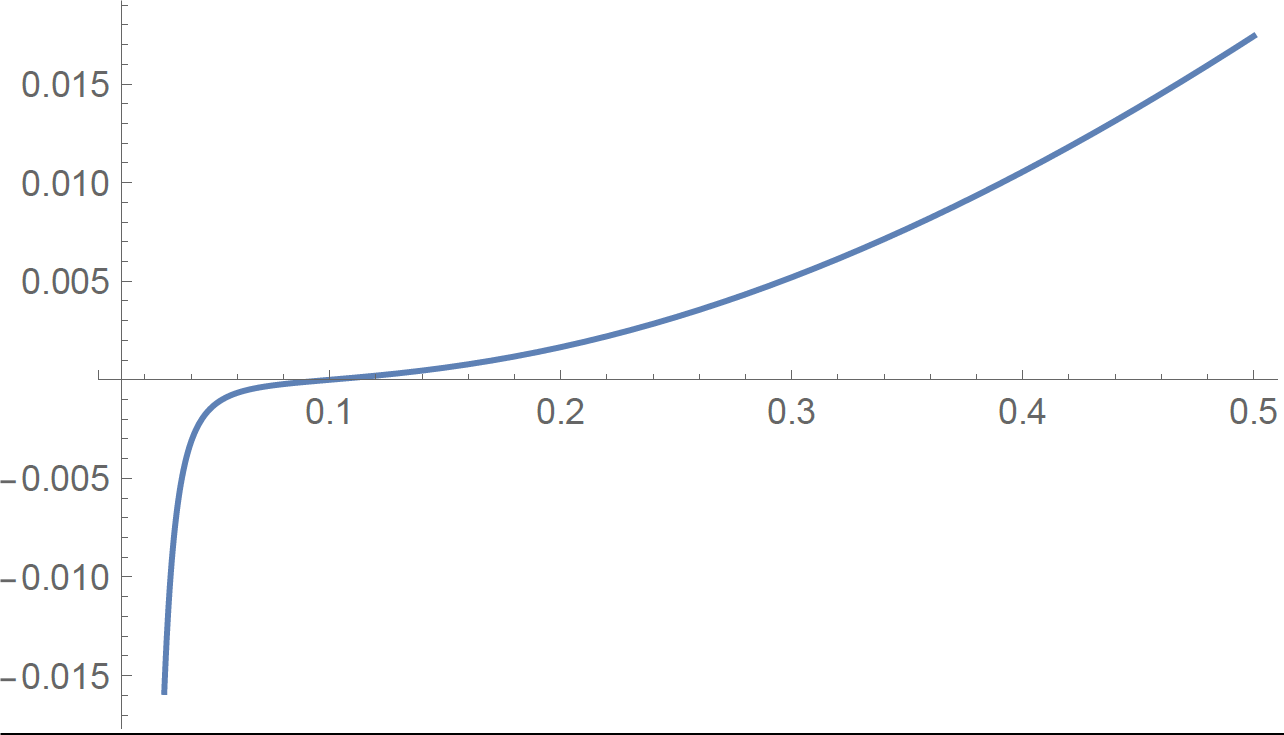}
\caption{$\ell(s)_{\text{high density}}$ dependence graph for $s_0=0.1, \ell_0=0, \dot\ell_0=0.01$}
\label{l(s>0)}
\end{figure}

Let us denote $s=\frac{\eta}{C}$. When $C>0$, then $s>0$ and equation \eqref{geoddev1} then becomes

\begin{eqnarray}
\frac{d^2\ell}{ds ^2}-\frac{\ell}{s^3}=0 \label{geoddevspos}\, ,
\end{eqnarray}
The solution to this equation is given by 
\begin{equation}
\ell(s)=\sqrt{s}\left[ g_1 I_1(2/\sqrt{s})+g_2K_1(2/\sqrt{s})\right]    \approx (s)^{3/4}(g_1 e^{\frac{2}{\sqrt{s}}}+ g_2 e^{-\frac{2}{\sqrt{s}}}),   \label{lss}
\end{equation}
where $I_1$ and $K_1$ are Bessel functions of the first and second kind, respectively, and $g_1$ and $g_2$ are constants and can be expressed in terms of $\ell$ and $\dot\ell$ for $s=s_0$:
\begin{eqnarray}
g_1=\frac{e^{-\frac{2}{\sqrt{s_0}}}}{2}\left\{ \ell_0 \left[ \frac{1}{s_0^{3/4}}+\frac{3}{4}\frac{1}{s_0^{1/4}}\right]-\dot\ell_0\,\, s_0^{3/4}\right\},\\
g_2=\frac{e^{\frac{2}{\sqrt{s_0}}}}{2}\left\{ \ell_0 \left[ \frac{1}{s_0^{3/4}}-\frac{3}{4}\frac{1}{s_0^{1/4}}\right]+\dot\ell_0\,\, s_0^{3/4}\right\}. 
\end{eqnarray}
Plugging $g_1$ and $g_2$ back, we find
\begin{equation}
\ell(s)_{\text{high density}}\approx \frac{1}{4}\left(\frac{s }{s_0}\right)^{3/4} \left(4 \ell_0 \cosh 2 \left(\frac{1}{\sqrt{s }}-\frac{1}{\sqrt{s_0}}\right)+\sqrt{s_0} (3\ell_0-4 s_0 \dot\ell_0) \sinh 2  \left(\frac{1}{\sqrt{s }}-\frac{1}{\sqrt{s_0}}\right)\right)
\end{equation}
The $\ell(s)$ dependance is given in Figure \ref{l(s>0)}. The energy density perturbation in this case, for the solution \eqref{phiksol} becomes
\begin{eqnarray} 
 \frac{\overline{\delta\varepsilon}}{\varepsilon_0}\approx \frac{4C_2}{x^3}=2\frac{C}{k^2\eta^3}=\frac{2}{k^2 C^2 s^3}\,.\label{endenspert}
 \end{eqnarray}
Thus, $s>0$ corresponds to the case of high density regions.

\subsection{$C<0$ case}

\begin{figure}
\centering
\includegraphics[scale=0.4]{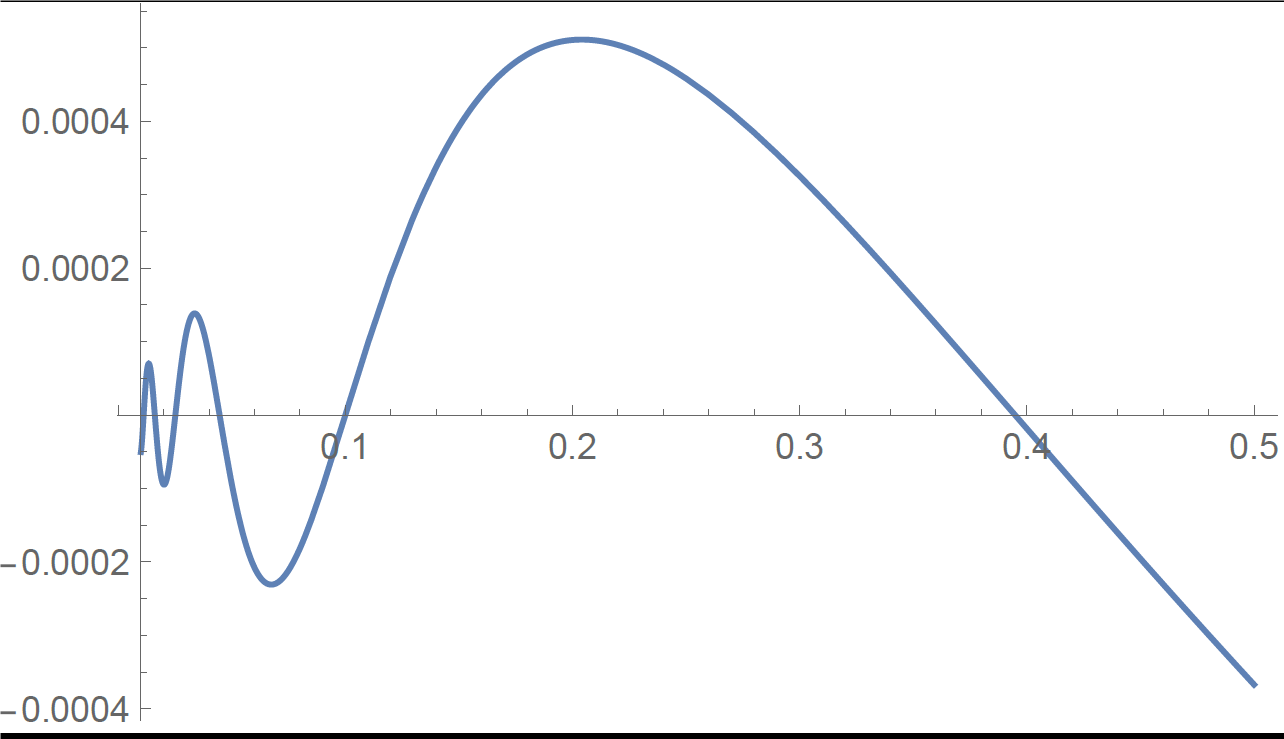}
\caption{$\ell(s)_{\text{low density}}$ dependence graph for $s_0=0.1, \ell_0=0, \dot\ell_0=0.01$}
\label{l(s<0)}
\end{figure}

In this case, let us introduce the variable $s=-\frac{\eta}{C}$. The minus sign ensures that $s$ grows along the time flow. The equation that we have to solve in this case becomes
\begin{eqnarray}
\frac{d^2\ell}{ds ^2}+\frac{\ell}{s^3}=0 \label{geoddevsneg}\, ,
\end{eqnarray}
In this case the solution is given by
\begin{eqnarray}
\ell(s)\approx s^{3/4}\left(g_1\cos\left(\frac{2}{\sqrt{s}}\right)+g_2\sin\left(\frac{2}{\sqrt{s}}\right)\right)
\end{eqnarray}
Again finding constants $g_1$ and $g_2$ and plugging them back, we find
\begin{equation}
\ell(s)_{\text{low density}}\approx \frac{1}{4}\left(\frac{s }{s_0}\right)^{3/4} \left(4 \ell_0 \cos 2 \left(\frac{1}{\sqrt{s }}-\frac{1}{\sqrt{s_0}}\right)+\sqrt{s_0} (3\ell_0-4 s_0 \dot\ell_0) \sin 2  \left(\frac{1}{\sqrt{s }}-\frac{1}{\sqrt{s_0}}\right)\right)
\end{equation}
The $\ell(s)$ dependance is given in Figure \ref{l(s<0)}. From equation \eqref{endenspert} it follows that the case $s<0$ corresponds to low density regions.

\newpage

\subsection{$C_2=0$ case}

For $C_2=0$, $\phi_k=\frac{C_1}{3}$ and the Jacobi equation \eqref{geodeq} takes the form
\begin{equation}
\frac{d^2\ell}{d \eta^2}-C_0\ell =0\, ,
\end{equation}
where $C_0=\frac{2}{3}k^2 C_1$. Its solution is given as
\begin{eqnarray}
\ell=g_1e^{\sqrt{C_0}\eta}+g_2 e^{-\sqrt{C_0}\eta}\, \label{elcnot},
\end{eqnarray}
where $g_1$ and $g_2$ are constants and can be expressed in terms of  $\ell_0=\ell(\eta_0)$, $\dot\ell_0=\dot\ell(\eta_0)$.
\begin{eqnarray}
g_1=\frac{e^{-2\sqrt{C_0}\eta_0}}{2}\left(\ell_0+\frac{\dot\ell_0}{2\sqrt{C_0}}\right),\qquad
g_2=\frac{e^{2\sqrt{C_0}\eta_0}}{2}\left(\ell_0-\frac{\dot\ell_0}{2\sqrt{C_0}}\right)\,.
\end{eqnarray}
Plugging the constants back in \eqref{elcnot}, we get
\begin{eqnarray}
\ell(\eta)=\ell_0 \cosh 2 \sqrt{C_0} (\eta -\eta_0)+\frac{\dot\ell_0} {2 \sqrt{C_0}}\sinh 2 \sqrt{C_0} (\eta -\eta_0).
\end{eqnarray}
The energy density perturbation for $C_2=0$ is 
\begin{eqnarray} 
 \frac{\overline{\delta\varepsilon}}{\varepsilon_0}\approx -\frac{2}{3}C_1= -\frac{C_0}{k^2}\, .
 \end{eqnarray}
 Hence $C_0>0$ corresponds to low density, while $C_0<0$ is for high density regions. Thus, we have
\begin{eqnarray}
\ell(\eta)_{\text{low density}}=\ell_0 \cosh 2 \sqrt{C_0} (\eta -\eta_0)+\frac{\dot\ell_0} {2 \sqrt{C_0}}\sinh 2 \sqrt{C_0} (\eta -\eta_0)
\end{eqnarray}
and for the high density regions
\begin{eqnarray}
\ell(\eta)_{\text{high density}}=\ell_0 \cos 2 \sqrt{-C_0} (\eta -\eta_0)+\frac{\dot\ell_0} {2 \sqrt{-C_0}}\sin 2 \sqrt{-C_0} (\eta -\eta_0).
\end{eqnarray}

\begin{figure}
\centering
\includegraphics[scale=0.4]{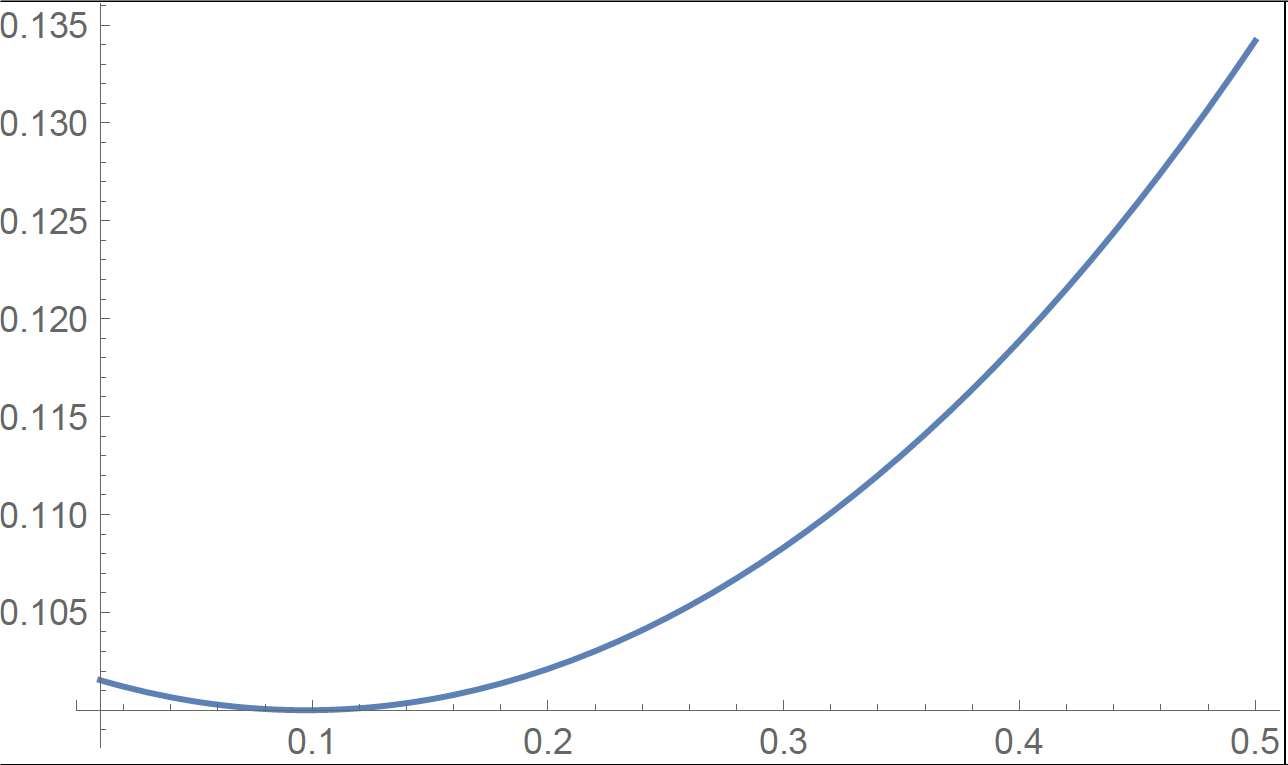}
\caption{$\ell(\eta)_{\text{low density}}$ dependence graph for $C_0=1, \eta_0=0.1, \ell_0=0.1, \dot\ell_0=0.001$}
\label{letavoid}
\end{figure}

\begin{figure}
\centering
\includegraphics[scale=0.4]{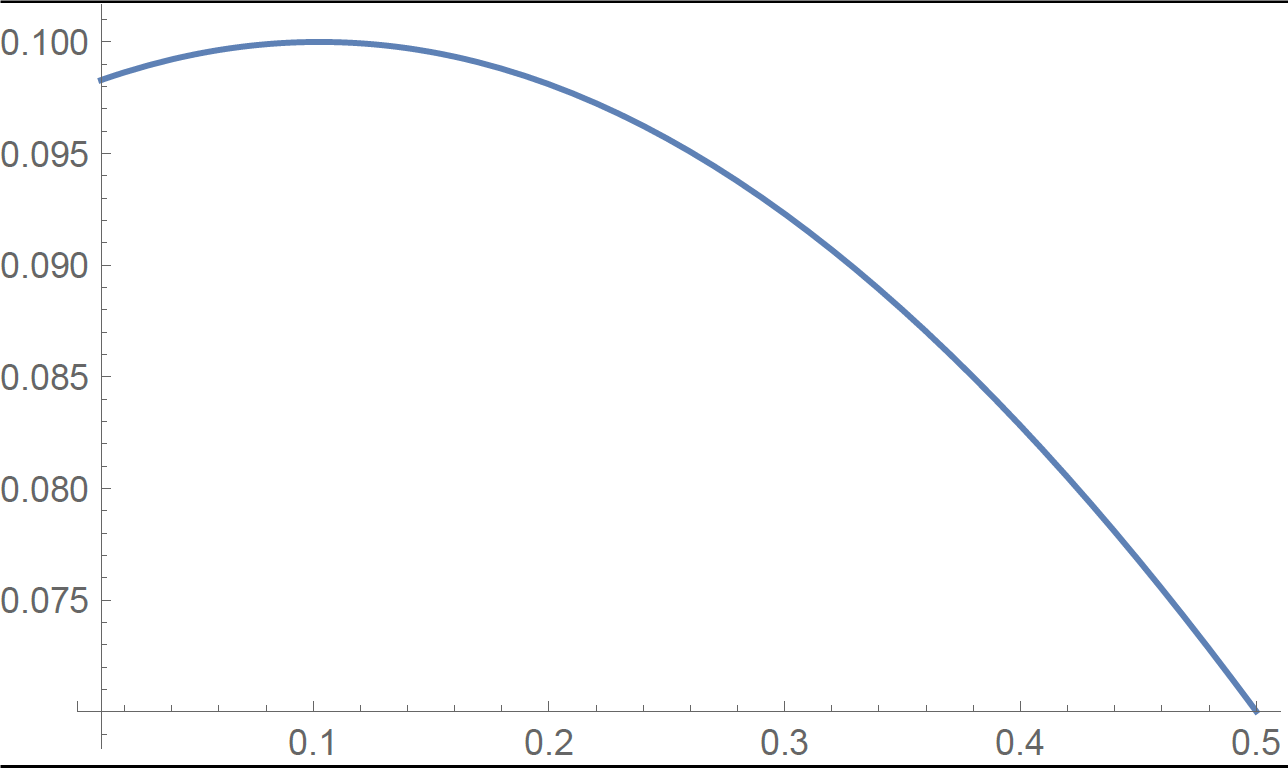}
\caption{$\ell(\eta)_{\text{high density}}$ dependence graph for $C_0=-1, \eta_0=0.1, \ell_0=0.1, \dot\ell_0=0.001$}
\label{letawall}
\end{figure}

Figures \ref{letavoid} and \ref{letawall} present the corresponding $\ell(\eta)$ dependencies, in the normalization of the formulae above. The figures clearly reveal the hyperbolicity of the geodesic flow due to the low density regions. 

\section{Conclusions}

We studied the hyperbolicity properties of the primordial gravitational waves due to the inhomogeneities in the radiation-dominated Universe. The behavior of geodesic flows in perturbed metric is studied analysing the solutions of the Jacobi equation of geodesic deviation enabling to trace the hyperbolicity features. The revealed hyperbolicity of the gravitational waves has to smear their original properties. 

Then the considered effect of smearing of primordial gravitational wave background is related to the following principal issue. It is expected that the primordial gravitational waves are preserving their properties, thus enabling direct comparison with the theoretical predictions regarding their origin in quantum fluctuations in very early Universe. However, in the case of smearing of the gravitational waves the recovering of the particular details of the inflationary phase, the scalar fields, etc, from the future experimental data would imply far more complicated procedure. If the smearing is occurring at the radiation-dominated phase, then it can be used for distinguishing of any gravitational background signal from a signal originated in matter-dominated phase, as e.g. due to supermassive black hole binaries believed to cause the nano-hertz signal detected by NANOGrav. 

The remarkable consequence of the smearing effect can be the flattening of the original spectrum of the gravitational wave background. The spectrum's flattening effect quantitatively has to be determined by such descriptors, as the Kolmogorov-Sinai (KS)-entropy \cite{Kolm} $h$ defining the mixing properties of geodesic flows and the decay of correlation functions \cite{Arn,Pol,GK1}. Namely, the correlation function  of the geodesic flow $\{f^\lambda\}$ defined as
\begin{equation}
   b_{A_1,A_2}({\lambda})=\int_{SM}A_1\circ f^{\lambda}\cdot A_2 d\mu
         -\int_{SM}A_1d\mu\int_{SM}A_2d\mu,
\end{equation}
on the unit tangent bundle $SM$ of Riemannian 3-manifold $M$ with negative constant curvature
decreases by exponential law for all functions $A_1,A_2\in L^2(SM)$ (see \cite{Pol} for details)
\begin{equation}
   \left|b_{A_1,A_2}(\lambda)\right|
      \leq c\cdot \left|b_{A_1,A_2}(0)\right|\cdot e^{-h\lambda} \ ,
\label{expmix}
\end{equation}
here $\mu$ is the Liouville measure, $\mu(SM)=1$. KS-entropy in this case has to be determined by the scales of the inhomogeneities in the radiation-dominated Universe \cite{GK2}. Thus, the considered effect of smearing can have direct consequences in the interpretation of the observational data on the primordial gravitational wave background, opening also a way to probe the density inhomogeneities in the radiation-dominated phase, linked to the cosmological tensions and modified gravity theories.

\section{Acknowledgment}
We thank the referee for valuable comments. M.S. is acknowledging the ANSEF grant 23AN:PS-astroth-2922.

\section{Data Availability Statement} 
Data sharing not applicable to this article as no datasets were generated or analysed during the current study.

\end{document}